# EasyEyes — Online hearing research using speakers calibrated by phones

Ivan Vican[1] (ivan@nanophonics.com)
Hugo De Moraes[2] (demoraeshugo@gmail.com)
Chongjun Liao[2] (cl6399@nyu.edu)
Nathnael H. Tsegaye[2] (nht249@nyu.edu)
William O'Gara[2] (weogara@gmail.com)
Jasper Inamoto[3] (jasper.inamoto@gmail.com)
Denis G. Pelli[2,4] (denis.pelli@nyu.edu)

1. *Nanophonics, Imotski, Croatia*
2. *Dept. of Psychology, New York University, New York, NY 10003, USA*
3. *Poly Prep Country Day School, Brooklyn, NY, USA*
4. *Center for Neural Science, New York University, New York, NY 10003, USA*

**ABSTRACT**

Hearing research requires a calibrated sound source, traditionally as lab equipment. Online research is quicker and more inclusive, but most participants lack calibration equipment and their sound sources are uncalibrated and diverse. This article explains how the open-source EasyEyes.app calibrates loudspeakers online. A library of smartphone-microphone profiles allows EasyEyes to use the participant's phone to calibrate their computer's loudspeaker in three minutes. Participants select their phone model, which is verified by screen size. Calibration employs the Novak et al. nonsynchronous maximum-length-sequence (MLS) algorithm. The computer's loudspeaker is corrected by convolving its input with the inverse of its impulse response. Researchers can contribute to the open-access library by calibrating phones with a measurement microphone. In the library, each profile is linked back to the profile used to produce it, back to the manufacturer profile of a measurement microphone. Correction accuracy is such that playing the flat-spectrum MLS through the corrected loudspeaker produces a nearly flat spectrum, with standard deviation less than 3 dB. A survey shows that a library of 94 phone models from major brands will support most participants in the USA (87%) and UK (80%). This method facilitates efficient and inclusive online hearing research.

## I. INTRODUCTION

*Online.* Perception research in psychology and neuroscience is increasingly done online because online recruitment is easy and reaches diverse populations (Mok et al. 2023; Peng et al. 2022; Woods et al. 2015). For online testing, the scientist implements the testing software as a web app that runs in the browser of the participant's computer, and typically uses a service, e.g. Amazon MTurk or Prolific, to recruit participants. Hearing research requires accurate sound generation, which demands calibration, traditionally with professional sound equipment. Several online testing platforms, including PsychoJS, lab.js, and Gorilla, facilitate psychophysical testing online, but none offer online frequency-response sound calibration. EasyEyes closes this gap, creating a phone profile library that combines with an established sound-calibration method to enable production of research-grade sound stimuli in online hearing research.

*ASA Task Force*. In May 2020, the Acoustical Society of America's Technical Committee on Psychological and Physiological Acoustics formed the Task Force on Remote Testing to evaluate remote testing methods and challenges for researchers (Peng et al. 2022). They identified two main approaches: take-home testing, where calibrated equipment is sent to participants, and web-based testing, where participants use their own equipment. They found 34 remote studies of hearing published from January 2020 to January 2022. Most (28 of 34) were web-based, and the rest were take-home or combined both methods. Notably, scientist-calibrated stimuli or hardware were used in most (3/4) take-home studies and none of the 28 web-based ones. Two of those studies asked participants to adjust a sound: in one to match a reference tone, in the other, to reach comfort. The committee concluded that, "By and large, remote testing methodologies offer greater flexibility and access to participants but poorer experimental control as compared to in-person data collection." They suggested using take-home equipment when tight stimulus control is required in remote testing. This paper reports a way to achieve relatively tight stimulus control in a web-based study.

The EasyEyes.app platform enables online research on visual crowding and related phenomena, including auditory informational masking (Kurzawski et al., 2023; Zhang et al., 2021).

*Overview*. This paper explains the nonsynchronous sound-system calibration algorithm, adapted from Novak et al. (2016). It plays a digital sequence through the speaker, uses the microphone to record the sound digitally, and uses the digital input and output to compute the impulse response. With a calibrated microphone, this algorithm calibrates a loudspeaker, and with a calibrated loudspeaker, it calibrates a microphone. We also explain how the profile library is built, and how it delivers microphone and loudspeaker profiles indexed by timestamp and by the phone or computer model. Finally we assess the accuracy of each step in the process.

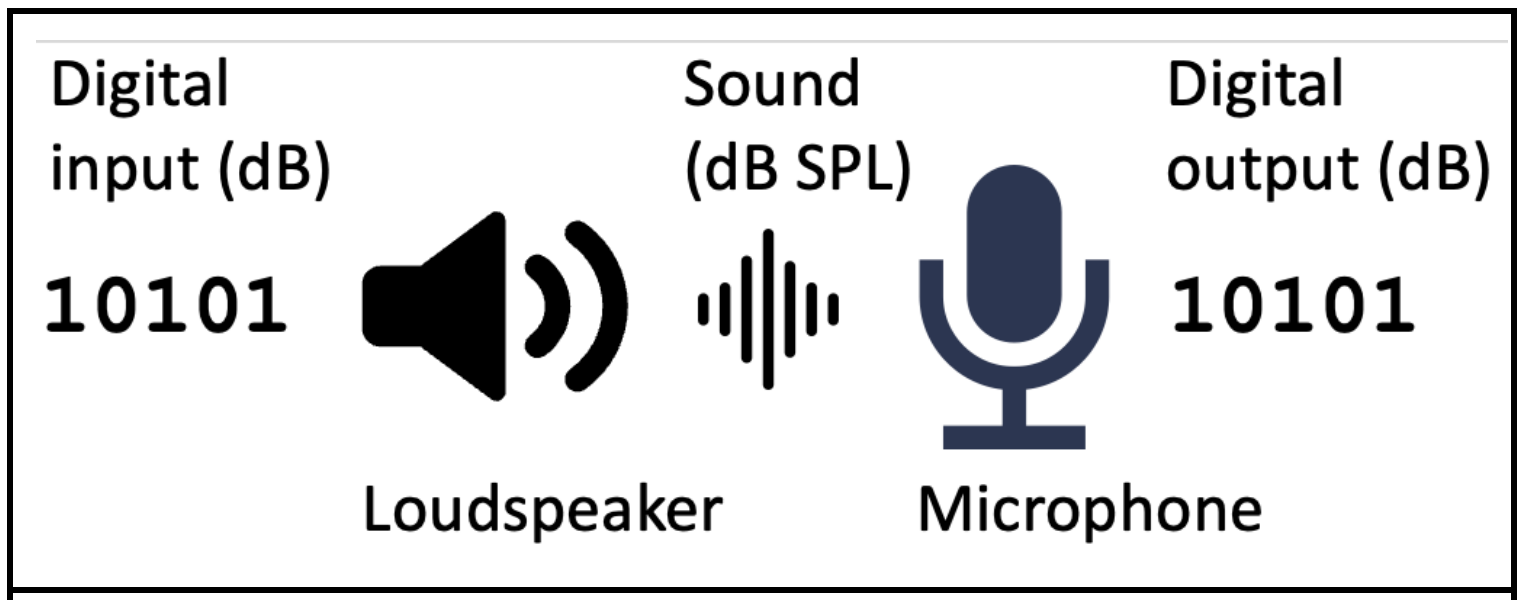

**Figure 1. Transducers.** The loudspeaker and microphone are transducers that convert numbers to sound or sound to numbers. Information flows from left to right.

*Loudspeakers, not headphones*. In the lab, hearing researchers typically deliver sound through calibrated headphones. Headphones typically have a better frequency response, higher volume, and better stereo separation than do the loudspeakers of consumer-grade computers. However, physical sound calibration of headphones requires an expensive coupler that online participants lack. Instead we focus on the computer loudspeakers. The scheme could be enhanced to handle stereo, but, for now, EasyEyes treats the two speakers as one.

Sound calibration requires a loudspeaker and a microphone. The essential elements are depicted in Fig. 1. A computer sends digital sound data to its loudspeaker, which produces sound, which is received by a microphone, which emits data. The microphone and loudspeaker are transducers converting back and forth between numbers and sound.

**Calibrating the loudspeaker and microphone**

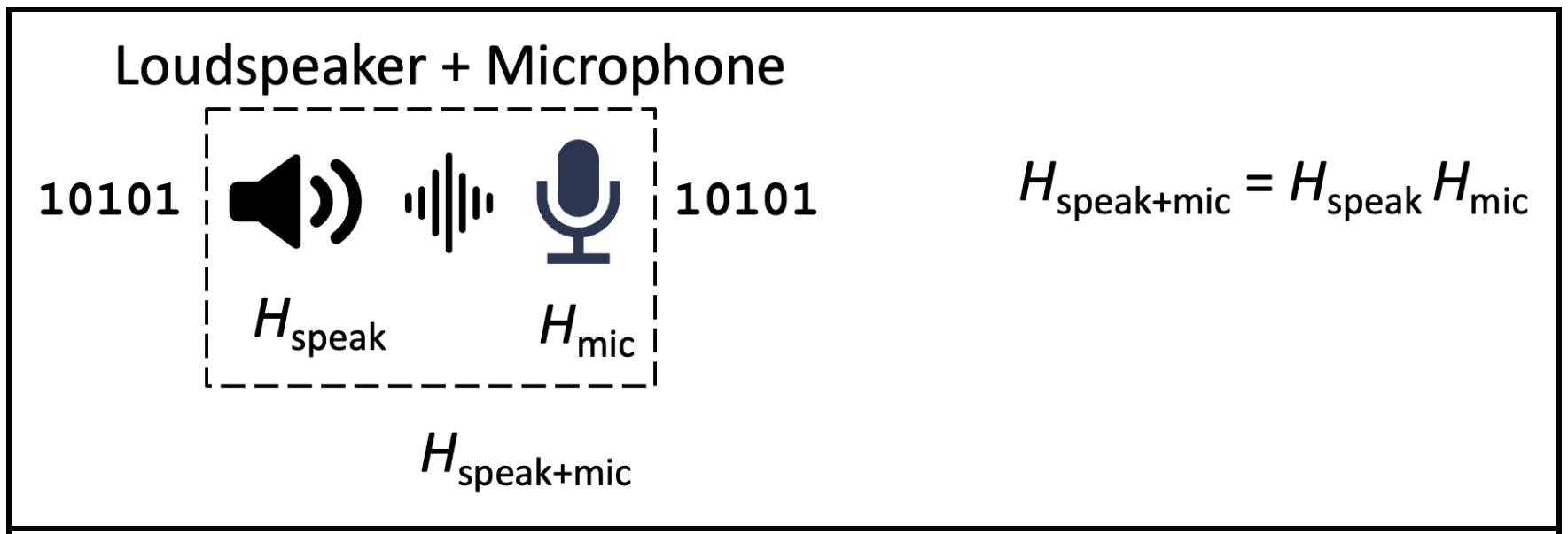

**Figure 2. Frequency response.** Each transducer has a frequency response, $H_{speak}$ or $H_{mic}$. In this paper, we typically know one and use it to measure the other. The Novak et al. (2016) method treats the combination of loudspeaker and microphone as a digital filter, and estimates its frequency response $H_{speak+mic}$. If we know the microphone's frequency response $H_{mic}$, we can calibrate the loudspeaker: Measure $H_{speak+mic}$, and compute $H_{speak} = H_{speak+mic} / H_{mic}$. If we know the loudspeaker's frequency response $H_{speak}$, we can calibrate the microphone: Measure $H_{speak+mic}$, and compute $H_{speak} = H_{speak+mic} / H_{mic}$.

Initially we treat the combination of loudspeaker and microphone as one linear system. Using a particular input, the Novak et al. (2016) algorithm computes the system's impulse response or its Fourier transform, the frequency transfer function $H$. If either the speaker or the microphone is already calibrated, then its frequency response can be factored out, leaving the frequency response of the other (Fig. 2). Thus one calibrated transducer can be used to calibrate another. The challenge is to calibrate the loudspeakers of an online computer without any professional sound-calibration equipment at the online site. A key part of the solution is the EasyEyes library of microphone profiles. Each microphone profile in the library contains the impulse response of a smartphone's microphone, indexed by the phone model. The library saves loudspeaker profiles as well, to document the calibration chain. Calibration of a loudspeaker requires a calibrated microphone, and calibration of a microphone requires a calibrated loudspeaker.
Thus each profile created by calibration has a "parent" profile used to create it. In the library, each profile is linked to its parent profile, which, in turn, is linked to its parent, and so on, until the linked list eventually ends with a manufacturer-calibrated measurement microphone profile (UMIK-1 or UMIK-2, from miniDSP, Hong Kong, costing US$79 and US$195 in 2024).

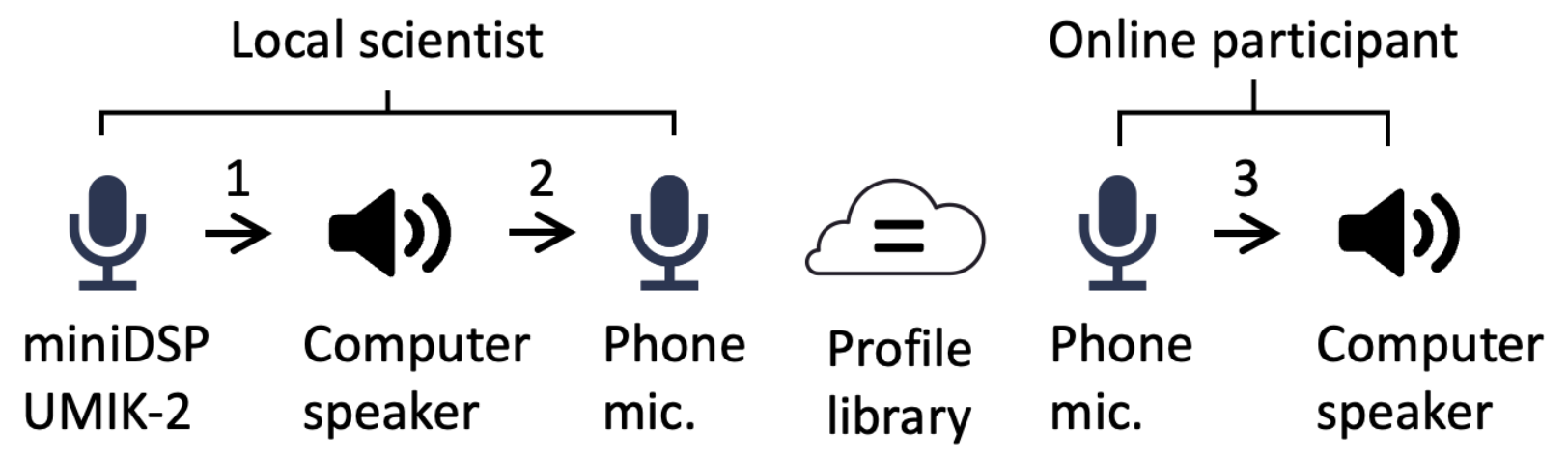

**Figure 3. Calibration chain.** The three calibration-chain links are numbered. Each arrow represents using the frequency response of the left transducer to calibrate (i.e. measure the frequency response) of the right transducer. The cloud copies the measured profile of one microphone to serve as the profile of another microphone with the same model number. The miniDSP UMIK-2 microphone is calibrated by the manufacturer. 1. It calibrates the speaker of a local computer, which 2. calibrates many phone microphones, one by one. All the phone microphone frequency responses are saved as profiles in the EasyEyes profile library in the cloud. Online, the appropriate profile is pulled from the library to 3. allow the participant's phone to calibrate the participant's loudspeaker. This use of the profile library relies on the working assumption that all phones with the same model name or number have practically the same frequency response, which we will assess as we grow the library.

Fig. 3 shows how EasyEyes manages calibration. A scientist uses a measurement microphone to calibrate a phone and saves the phone's profile in the library. At some future date a participant with that model phone matches the library and uses that profile to calibrate their loudspeaker.

*Traceability*. Metrological traceability requires a "documented unbroken chain of calibrations, each contributing to the measurement uncertainty" (VIM, 2007). Thus the EasyEyes profile library includes both loudspeaker and microphone profiles, and each profile is linked to the "parent" profile that was used to produce it (e.g. a microphone profile is used to calibrate a speaker). Each linked list ends with a measurement microphone. Uncertainty is assessed by the accuracy of correction. After estimating the impulse response of the next transducer, EasyEyes computes the inverse impulse response, which it convolves with the MLS. This digital signal is bandlimited to not exceed the power of the unfiltered MLS and played and recorded. Success in correction is assessed by the SD (in dB) of the spectral sound level across the corrected band, typically achieving an SD less than 3 dB over 100 to 10,000 Hz. When the SD exceeds 3 dB, the calibration is repeated.

In Results, we report ten-fold repeated calibrations, to assess the standard deviation of repeated measurement.

# II. METHODS

## A. Calibration management

### *1. EasyEyes online platform*

EasyEyes consists of three one-page web apps. One is a compiler used by the scientist to create the second, which is an experiment that runs in the browser of the participant's computer. The third web app runs on the participant's phone. The scientist controls everything simply by listing parameters with assigned values in a spreadsheet. The parameters are documented in a glossary (https://easyeyes.app/ select Docs : Parameter Glossary). The Methods below refer to many options set by the scientist. Each is a parameter assigned a value in the spreadsheet.

The EasyEyes scientist page is in English. The two participant pages use Google Translate to translate all instructional text into practically any language selected by the scientist or participant, at the scientist's discretion. Right to left languages like Arabic and Hebrew are supported.

*"Needs" page.* The computer web app first presents a Needs page, a gateway that tests whether the study's hardware and software requirements are met, especially whether the participant's phone is supported by the EasyEyes profile library. The needs may also include the computer's number of CPU cores and which browser. To minimize the chances of turning away participants at the Needs page, all the requirements that are enforced by the Needs page are advertised in the study Description that participants read before choosing the study. For studies that need a profile-library-supported phone, the Description will list the currently supported phone brands. Once the Needs page approves, participants proceed to the Consent page, where they can review and agree to an IRB-approved consent form.

To help check the phone's eligibility, a QR code is shown on the computer screen, which, when decoded by the mobile camera and tapped by the participant, brings up the mobile web page. The mobile web page queries the mobile phone, including the screen size in pixels, and communicates results back to the computer web page. The computer web page provides brand-specific instructions to guide the participant to find and enter the model name and number of the phone and computer.

### *2. Beyond the lab, challenges of online testing*

*Coping with consumer-grade equipment*. Hearing researchers stock their labs with professional sound equipment. EasyEyes relies on consumer phones and computers. The obvious difference between professional and consumer sound equipment is that professional equipment is linear and costs more. Less obvious is that the consumer-grade devices are more complicated, incorporating clever nonlinear compromises that compensate for severe hardware limitations, but complicate calibration. Web audio, by default, does echo cancellation, noise suppression, and automatic gain control. The details don't concern us here because any of these will ruin linear system calibration. Fortunately there are web audio calls to turn them off.

*Avoiding dynamic range compression.* Another consumer-grade issue is that linear systems analysis requires linearity, but computer loudspeakers typically feature a built-in nonlinearity called "dynamic range compression", which cannot be disabled, that allows a wide-dynamic-range recording to be played without clipping on small-dynamic-range loudspeakers (Giannoulis et al., 2012). Soft sounds are played linearly and loud sounds are compressed. EasyEyes respects this limit by calibrating at the loudest linear sound level. EasyEyes measures the system gain at 1 kHz over a wide input range (-50 to -3.1 dB), fits a dynamic-range-compression model (which typically fits well, rms error < 0.5 dB), which has a breakpoint where the dB-out vs dB-in slope drops from 1 to 0.5. Linearity is also assessed by total harmonic distortion (THD). Typically THD tracks the log-log slope so that when the slope is 1, the THD is quite low, around 0.1%, and when the log-log slope is 0.5, the THD is higher, around 0.5%. In the next step, EasyEyes measures the frequency response at a sound level 14 dB below the breakpoint.

*Preventing sleep*. Smartphones save battery power by going to sleep quickly, which spoils any ongoing sound recording. Fortunately the widely supported Wake Lock web API keeps the phone awake (https://developer.mozilla.org/en-US/docs/Web/API/Screen_Wake_Lock_API).

*Web apps are not trusted*. In the lab, the scientist's software runs as a trusted app on their computer, with full access to all system information. Online, the scientist's software runs as a web app in the participant's browser. Web apps are not allowed to set or read the sound volume, so EasyEyes asks the participant to do that for us. Furthermore, modern browsers do not trust web apps and deny access to most information that might identify the participant, including the computer's or phone's model name. Here too, EasyEyes asks the participant for help, as explained in the next section.

### *3. Identifying the phone model*

Identifying the correct phone model is crucial. Web privacy rules prevent browsers from identifying the phone model so participants manually indicate their phone's model name and number. The model name and number are requested, one after the other, and checked for consistency with each other, and the browser-reported screen size (in pixels). This helps verify the phone model because each model year tends to have a larger screen. Thus the 41 iPhone models have 13 different screen sizes.

EasyEyes looks up the phone in the profile library by model name, number, and screen size. If a 3-way match is found, the participant is congratulated, and the experiment proceeds. Partial matches are treated as failure to match. When matching fails, EasyEyes expresses regret and encourages the participant to try with another phone. We anticipate that this combination of two user-specified identifiers (model name and number) and one electronically queried detail (screen size) will identify phones quite reliably.

*Growing the library, brand by brand.* We will minimize runtime rejection of unsupported phones by providing full support of a few phone brands and alerting potential participants that only these brands are supported in the study description they read while deciding whether to participate. As we build the profile library, we prioritize phone brands by popularity: first Apple, then Samsung, Google, etc. Our survey, presented in Results, finds that the three most popular brands suffice to support the phones of 87% of participants in the USA and 80% in the UK. We aim to fully support each brand before moving on to the next.

### *4. Identifying the computer's sound output device*

*Internal selection of sound output device.* Computers offer users a choice among available sound generators, including headphones and loudspeakers. Researchers need to know which device is used by the experiment. Perceptual tests can distinguish loudspeakers from headphones (Woods et al. 2017). EasyEyes uses the Web Media Devices API to get the human-readable name of the selected sound output, e.g. "MacBook Pro Speakers (Built-in)" (google "MDN Media Streams"). On a computer (desktop or laptop), this typically reveals the brand and type of the selected sound input and output, though not the model. On mobile devices (phones) it is more restricted and typically reveals only which sound devices are available, not which is selected for use.

### *5. Profile library*

Calibration of a microphone or speaker yields an impulse response, which is converted by FFT to a frequency response and saved as a sound profile (gain and phase at each frequency, from 0 to 20 kHz) in the EasyEyes profile library. (The upper frequency limit is selected by the scientist.) The library holds microphone profiles indexed by smartphone model and loudspeaker profiles indexed by computer model. They are also indexed by timestamp, which is used to link each profile to its parent profile.

*The library is growing.* A local used-phone store (smartphonenycfix.com) has thousands of used phones and allows in-store testing of their used phones, paying only a small fee per phone.
The plan is to first calibrate as many as possible of the 41 Apple iPhone models, the most popular brand, and then Samsung, Google, and Huawei. Once all the models of a brand are in the library, potential participants can be told EasyEyes supports phones of that brand.

*Contributing to the library.* Using the EasyEyes.app, any scientist with a computer and a measurement microphone (e.g. the US$79 UMIK-1 from miniDSP, Hong Kong) can calibrate phones and contribute their profiles to the EasyEyes open-access library. Calibration takes three minutes per phone. Each microphone profile is "signed" with the responsible scientist's email. Any scientist planning an experiment specifies which signatures are approved in matching the phones of their participants.

*UMIK-1 vs. UMIK-2 microphone.* The UMIK-2 has lower noise and offers more sampling frequencies, but the two microphones are equivalent when used with EasyEyes for sound calibration in a typically noisy environment. The UMIK-2 is exactly 0.5 inch in diameter and is thus compatible with calibrated sound level sources, like the Reed R8090, that fit tightly to produce precisely 94 dB SPL. The UMIK-1 has a smaller diameter, preventing a tight fit, making it incompatible with such 0.5 inch calibrators, but is compatible with the Center 326 regulated sound level calibrator, which doesn't require a tight fit (Center Technology Corp., Taiwan, www.centertek.com). Using the REW software (www.roomeqwizard.com) and the Reed R8090 and Center 326 calibrators, our UMIK-1 and UMIK-2 microphones correctly read the sound level of the 1 kHz 94 dB SPL tone within ±0.5 dB.

### *6. Calibration instructions*

If the phone is supported by the profile library, then the participant is congratulated and guided in using the phone to calibrate the loudspeaker, and the new loudspeaker profile is used to correct it, as explained below. Failing a phone match, the participant is invited to try another phone. Failing that, EasyEyes says “sorry” and asks the participant to “return” the study to Prolific, i.e. quit.

*General microphone placement*. Ideally we would measure the free-field sound level where the participant’s ear will later be. In practice, for compliance, the instructions must be easy to follow. Our microphone placement instructions put the measurement microphone not far from where the participant’s ear would be. The phone placement is lower, because no tripod is handy, but this is matched between creation and use of the phone’s profile. Computer speakers may, to some extent, beam their sound energy toward the participant’s face so we attempt to place the phone microphone in that beam. If the participant places their computer farther than the computer used in calibration, then both the phone and the participant’s ears will experience a reduced sound level. These matched effects tend to cancel, conserving the accuracy of the estimated the sound field near the participant’s ear.

*Placement of the measurement microphone.* The measurement microphone is used solely by a scientist who will calibrate phone microphones. It is placed on the table between the scientist and the computer. EasyEyes uses the microphone’s 90° profile from the manufacturer, and tells the scientist to point the microphone up toward the ceiling. The microphone is in its tripod, with the microphone’s bottom flush with the bottom of the tripod’s holding clip. This places the microphone’s tip 26 cm above the table, and roughly on the scientist’s line of sight to the middle of the keyboard. (The line of sight is a rough approximation for the line of hearing, which is less easily discerned by the scientist and participant.) Laptop speakers usually reside in the keyboard. With a MacBook Pro, the microphone tip is 32 cm from the center of the keyboard surface. The owner of that computer typically sits with his ears 54 cm from the center of the keyboard.

*Phone placement.* Once the phone is in communication with the computer (by Web RTC), the participant is asked to remove any protective case from the phone, and:

> “REST PHONE ON A TALL DRINKING GLASS (or coffee mug or soda/beer can), between you and your keyboard, with the microphone (bottom) edge facing your computer. Laptop speakers are in the keyboard; place your phone so that its microphone is in your line of sight to the keyboard. Desktop speakers are behind a grill; point your phone at the grill.”

This is intended to place the microphone near the participant, about 15 cm above the table, facing the loudspeakers. With a MacBook Pro, this places the phone’s microphone 18 cm from the center of the keyboard surface.

Scientists and participants tend to hurry to get the calibration started, so, once the calibration begins, EasyEyes presents a reminder:

> "⚠️ CAUTION. Did you remember to:
> ☐ maximize the computer sound volume?
> ☐ remove the phone's case?
> If not, hit RESTART, fix the problem, and PROCEED."

## B. Signal processing for calibration

### *1. 1000 Hz sine calibration*

*Gain and Total Harmonic Distortion (THD).* The input is a sine at frequency $f$ = 1000 Hz. Since the microphone and computer clocks are nonsynchronous, the recorded frequency $f'$ is slightly different, and this is estimated by finding the argmax of the FFT spectrum of the recording. For a MacBook Pro and an iPhone 12 Pro, over thirty 1 s recordings, the ratio $f'/f$ has a mean near 1 and a standard deviation around 0.00002. To reject background noise, EasyEyes cross-correlates the recording with a sine of frequency $f'$ to measure power solely at that frequency. Sine and cosine correlations are combined to make a phase-independent power measurement. Power is measured at frequencies $f'$, $2f'$, $3f'$, $4f'$, $5f'$, and $6f'$. Define $P_i$ as power at frequency $if'$. The analysis reports fundamental power $P_1$ at $f'$ and the Total Harmonic Distortion, $THD$, which is the square root of the ratio of total power of the 2nd to 6th harmonics divided by power of the fundamental,

$$THD = \sqrt{\frac{P_2 + P_3 + P_4 + P_5 + P_6}{P_1}} \qquad (1)$$

(Wiki THD, https://en.wikipedia.org/wiki/Total_harmonic_distortion). In our measurements THD typically ranges from 0.1% at low levels to around 0.5% at high levels. Plots of 1000 Hz gain and THD appear in the Results section.

***2. Maximum Length Sequence (MLS) calibration***

A Maximum Length Sequence (MLS) is used in a popular method for measuring the impulse response of an acoustic system (Stan, Embrechts, and Archambeau, 2002). A pseudorandom zero-mean binary sequence is played through the system and recorded. Analysis of the sequence and its distorted recording yields a linear model of the system that captures its acoustic parameters, including frequency response and reverberation. Even though the method tolerates impulsive interference and has a high signal-to-noise ratio, the impulse response is seriously distorted by violations of the assumed linearity of the system and synchronicity of the loudspeaker and microphone clocks (Nielsen, 1997). The latter is particularly pertinent to our case, since the computer and phone have independent clocks. Fortunately, Novak, Rund, & Honzík (2016) showed that the difference in clock frequency can be measured and corrected, yielding performance equivalent to that attained with synchronous clocks.

***3. Playing and recording the MLS***

EasyEyes generates a Maximum Length Sequence with a period length close to whatever duration the scientist specifies ( *_calibrateSoundBurstSec*, default 1 sec). The scientist specifies how many periods of playing will be analyzed ( *_calibrateSoundBurstRepeats*, default 4). The algorithm requires analysis of at least 2 periods. Following good practice, EasyEyes plays, continuously, first an extra period, then the requested number of periods for analysis, and finally, without any break, an extra 10% to allow for different sampling frequencies between the loudspeaker and microphone. The initial period is recorded but not analyzed, and is intended to give time for the system to settle into a steady state. To avoid clamping, the amplitude of the MLS should not exceed the range [-1,1], and is set by *_calibrateSoundBurstDb* (default -34 dB), producing a digital amplitude of ±10^(*_calibrateSoundBurstDb* / 20). The sampling rate is set by Web Audio to the available frequency whose log is nearest to the log of whatever sampling frequency the scientist requests, *_calibrateSoundSamplingDesiredHz* (default 48000).

***4. Computing the impulse response of the system: speaker, room, and microphone***

Before calculating the impulse response, we focus on correcting the difference in sampling rate between the computer loudspeaker and the phone. The sound is played and recorded for the desired number of sample periods, but the difference in sampling frequencies of the two devices will result in recording a different number of samples (in the phone) than were played (in the speaker). Even a small discrepancy can drastically affect the estimated impulse response.

Following Novak et al. (2016), EasyEyes estimates the period of the repeated-MLS recording by measuring the distance between peaks in the autocorrelation function of the recording. Measuring the distance between furthest peaks increases the precision of the estimated period. Next, EasyEyes stretches time to adjust the period length in the recording to match the original MLS period. This is done by applying the FFT to the several periods of the recording, and adding or removing elements at the end of the Discrete Fourier Transform (DFT) to adjust the length to match that of the original MLS. Finally, EasyEyes calculates the impulse response as the circular cross-correlation in the frequency domain of the recorded (and truncated) MLS with the original MLS. Let *Y* be the DFT of the several recorded MLS periods, *X* be the DFT of the original MLS, and *X** be the complex conjugate of *X*. The impulse response *IR* is the Fast Fourier Transform of the product of *Y* and *X**,

$$IR[n] = \frac{1}{L}\sum_{k=0}^{L-1} Y[k]X^*[k]e^{j\frac{2\pi kn}{L}}, n \in [0, L-1] \quad (2)$$

where *L* is the length of both *X* and *Y*, and *n* indexes the samples of the calculated impulse response. As noted above, the first recorded MLS period is ignored, to allow the system to settle, and the requested number of periods (more than 1) is analyzed.

***5. Discounting one transducer to recover the impulse response of the other***

The Novak et al. (2016) algorithm yields the impulse response of the combination of loudspeaker and microphone (Fig. 2). If the impulse response of either is known, it can be discounted to yield the impulse response of the other.

The impulse response measurement includes room acoustics in the “loudspeaker” impulse response, so “loudspeaker correction”, below, will correct both the loudspeaker and the room, provided the scientist specifies a sufficiently prolonged impulse response to include the room’s echoes.

***6. The impulse response (IR)***

The calculated impulse response is the length of the original MLS sequence (_calibrateSoundBurstSec, default 1 s). EasyEyes allows the scientist to limit the length of the impulse response (_calibrateSoundIRSec, default 0.2 s) and the inverse impulse response (_calibrateSoundIIRSec, default 0.2 s). The specified duration is centered on the peak. EasyEyes zeroes phase when restricting the impulse response length. Given the speed of sound (343 m/s), cutting off the impulse response at 0.2 s removes the first echo from a wall 35 m away, and the second echo from the walls of a 35 m room.

EasyEyes calculates the magnitude frequency spectrum of the impulse response, zeroes phase, and then applies the inverse DFT to get the signal back in the time domain and centers the waveform at time zero. The scientist specifies the desired length of the impulse response (default 0.2 s), which is multiplied by a Hann window.

### *7. Calculating the inverse impulse response (IIR)*

We ignore phase, and invert just the magnitude frequency spectrum, because phase differences in a room environment are mostly inaudible and might provoke correction issues (Norcross et al., 2004; also see Bank 2013 and Karjalainen and Paatero 2006). We generate a linear-phase impulse response filter, which we plan to upgrade to minimum-phase..

The frequency spectrum of the inverse $H_{inv}$ is calculated as

$$H_{inv}[k] = \frac{H^*[k]}{H^*[k]H[k]} \ , k \in [0, L-1] \tag{3}$$

where *H* represents the magnitude frequency spectrum of the truncated impulse response, and *H** represents the complex conjugate of *H*.

The last part of the procedure is to return the inverse of the impulse response in the time domain. We follow the procedure for impulse response truncation and apply the same approach here.

### *8. Correcting the loudspeaker*

We correct the loudspeaker output by convolving its digital input with the inverse impulse response signal:

$$y_{filt}[n] = \sum_{m=0}^{L-1} h_{inv}[m]x[n-m] \ , n \in [0, N+L-2] \tag{4}$$

where $y_{\text{filt}}$ is the output signal with applied correction, $h_{\text{inv}}$ is the inverse impulse response of length *L* that is used as a correcting filter, and *x* is the input recording with length *N*. We compensate for the filter delay by cutting out the first *N*/2 samples from the output file.

## *9. Checking the correction*

We again play and record the MLS digital sound, now convolved with the IIR, and plot the spectrum of the recording. If the correction is successful then the recorded spectrum will be flat, except for added background noise.

## *10. Monitoring power fluctuation*

When the participant has a supported phone, EasyEyes uses it to calibrate the participant's loudspeaker. We don't know why, but we discover that despite the steady signal power (sine or MLS) over each test interval, occasionally the recorded sound power fluctuates wildly (±40 dB) during a recording, contrary to the assumed stationarity. We check for that by monitoring the power while each sound plays. This measurement is just recorded power. There is no cross-correlation to reject background noise. It is typically quite flat, as seen in Figs. 5 and 6, but we occasionally see huge 40 dB swings up and down that are not reproduced when we repeat the measurement. When the power fluctuates, all the reported values become nonsense. The SD of power over time is a convenient scale for detecting this problem, and triggering recalibration. In Figs. 4 and 5 most of the curves have a standard deviation below 1 dB. The two faintest recordings of the 1000 Hz sine wave fluctuate more because they include background noise (street traffic). Again, these plots are crude, simply monitoring total power, but the calibration of gain and frequency response strongly rejects background noise by cross-correlation with the sine or MLS signal.

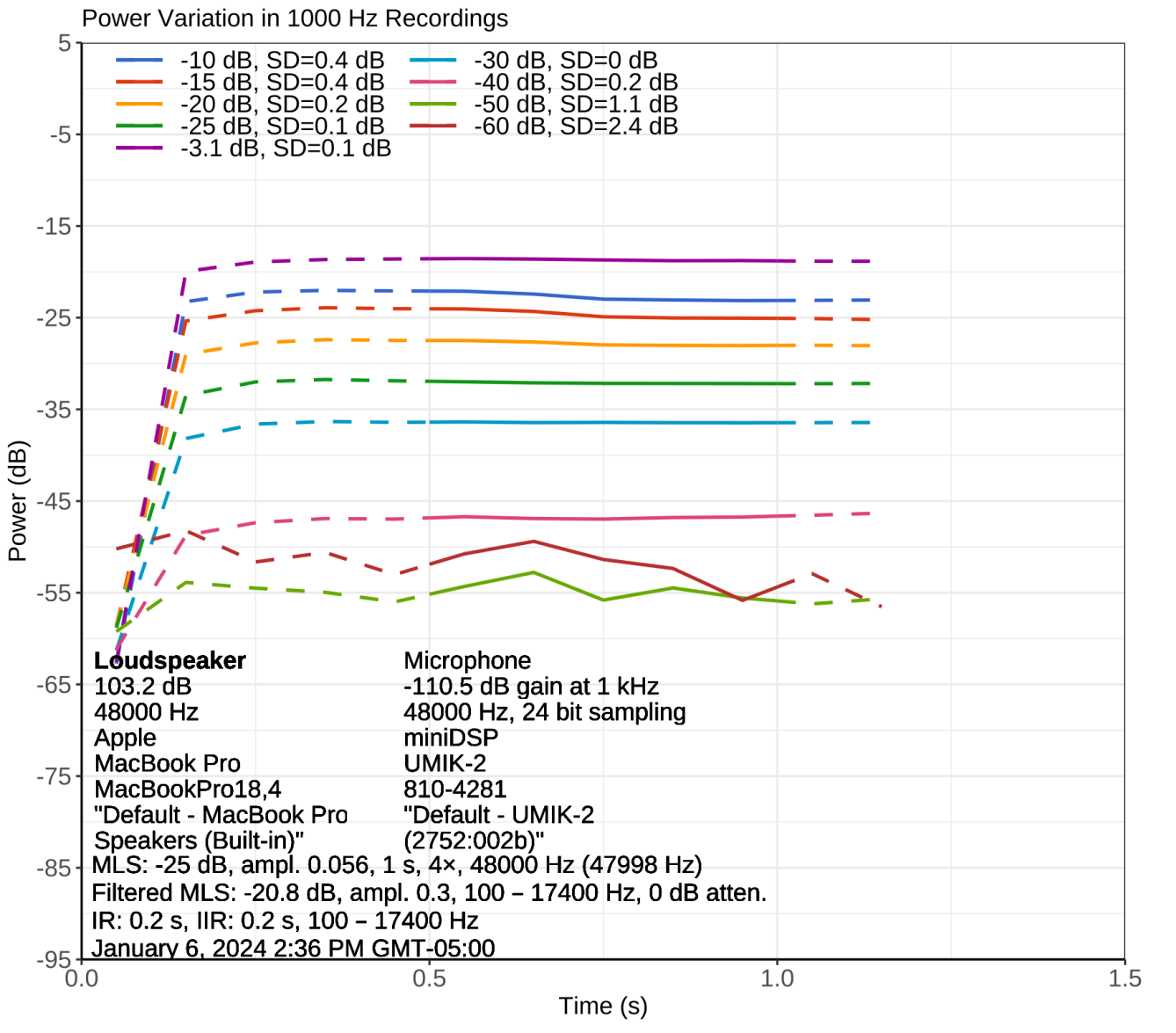

**Figure 4. Power variation during 1000 Hz recording.**

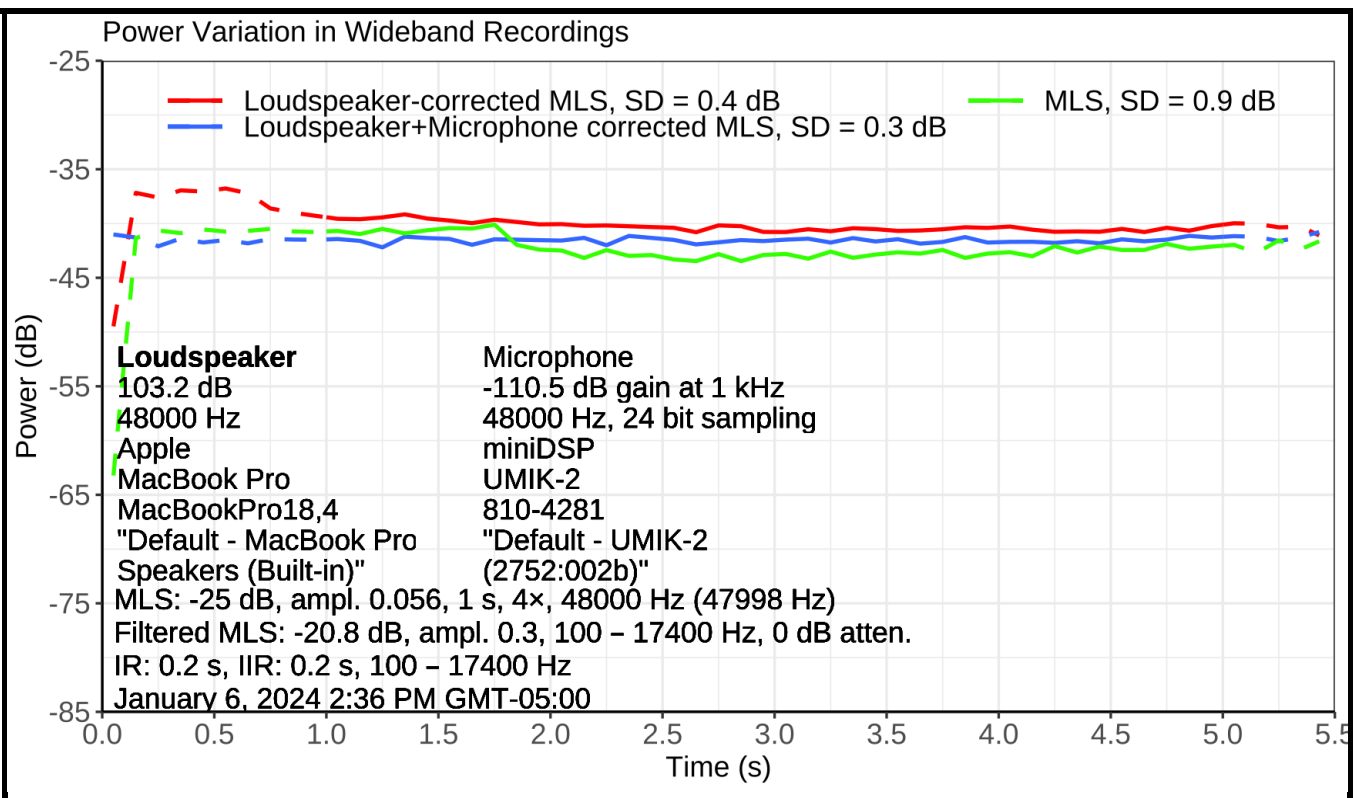

**Figure 5. Power variation during MLS recordings.**

***11. Summary of signal processing***

Once begun, the signal processing (2.2.1 to 2.2.10) is automatic and takes 2 minutes. (The three minutes to calibrate a phone include another minute to identify and set up the phone.) A graphic progress meter allows the participant to take a break. The final check (2.2.9) measures the accuracy of the loudspeaker correction. The passband of that check is set dynamically to keep power within the loudspeaker's linear range. The lower cut off is set by the scientist (calibrateSoundMinHz, typically 100 Hz), and the upper cut off (typically 3 to 20 kHz) passes as much power as the scientist allows (calibrateSoundBurstDb+calibrateSoundBurstFilteredExtraDb, typically -29 dB). In our experiments, we reject and redo any correction that yields a corrected spectrum in which the SD of power spectral density (in dB) across the passband exceeds 3 dB.

## III. RESULTS

### A. Phone-model survey

We surveyed 199 participants in the USA and 188 in the UK, identifying the participant's model of phone (Table 1). In interpreting the table, we assume that two phones with the same model name will have identical microphone acoustics. As we build the microphone profile library, we will test this assumption by testing at least two phones of each model.

| Phone brand | # | # USA | # UK | Cumulative proportion of participants | Cumulative proportion of USA participants | Cumulative proportion of UK participants | Phone models | Cumulative models | Proportion of participants per model | Operating system |
|---|---|---|---|---|---|---|---|---|---|---|
| Apple | 185 | 102 | 83 | 0.48 | 0.51 | 0.44 | 31 | 31 | 0.015 | iOS |
| Samsung | 116 | 59 | 57 | 0.78 | 0.81 | 0.74 | 52 | 83 | 0.006 | Android |
| Google | 24 | 13 | 11 | 0.84 | 0.87 | 0.80 | 12 | 95 | 0.005 | Android |
| Motorola | 17 | 12 | 5 | 0.88 | 0.93 | 0.83 | 14 | 109 | 0.003 | Android |
| Xiaomi | 9 | 2 | 7 | 0.91 | 0.94 | 0.87 | 8 | 117 | 0.003 | Android |
| Huawei | 8 | 1 | 7 | 0.93 | 0.95 | 0.90 | 7 | 124 | 0.003 | Android |
| OnePlus | 7 | 4 | 3 | 0.95 | 0.97 | 0.92 | 6 | 130 | 0.003 | Android |
| LG | 4 | 3 | 1 | 0.96 | 0.98 | 0.93 | 4 | 134 | 0.003 | Android |
| Oppo | 3 | 0 | 3 | 0.96 | 0.98 | 0.94 | 3 | 137 | 0.003 | Android |
| Honor | 2 | 0 | 2 | 0.97 | 0.98 | 0.95 | 2 | 139 | 0.003 | Android |
| Nothing | 2 | 0 | 2 | 0.97 | 0.98 | 0.96 | 1 | 140 | 0.005 | Android |
| Sony | 2 | 0 | 2 | 0.98 | 0.98 | 0.97 | 2 | 142 | 0.003 | Android |
| CUBOT | 1 | 0 | 1 | 0.98 | 0.98 | 0.98 | 1 | 143 | 0.003 | Android |
| IMO | 1 | 0 | 1 | 0.98 | 0.98 | 0.98 | 1 | 144 | 0.003 | Android |
| Infinix | 1 | 0 | 1 | 0.99 | 0.98 | 0.99 | 1 | 145 | 0.003 | Android |
| Nokia | 1 | 0 | 1 | 0.99 | 0.98 | 0.99 | 1 | 146 | 0.003 | Android |
| RCA | 1 | 1 | 0 | 0.99 | 0.99 | 0.99 | 1 | 147 | 0.003 | Android |
| T-Mobile | 1 | 1 | 0 | 0.99 | 0.99 | 0.99 | 1 | 148 | 0.003 | Android |
| TCL USA | 1 | 1 | 0 | 1.00 | 1.00 | 0.99 | 1 | 149 | 0.003 | Android |
| Ulefone | 1 | 0 | 1 | 1.00 | 1.00 | 1.00 | 1 | 150 | 0.003 | Android |
| TOTAL | 387 | 199 | 188 | | | | | | | |

**TABLE 1. Phone survey of 387 participants**. This table helps estimate the required profile library size to achieve a high probability of including a random participant's phone. Through the Prolific online recruitment service, we surveyed 387 participants, 199 in the USA and 188 in the UK, all fluent in English, and each with a computer whose CPU has at least 4 cores. The inclusion criteria are those of our current reading research. In fact, EasyEyes is multilingual, capable of automatic translation of its instructions into most languages. Of the 387 participants surveyed, 48% used Apple iPhones, with 31 different model names. The table shows that creating 95 profiles would suffice to support the Apple, Samsung, and Google phones surveyed (88% of participants). Three brands, Apple, Samsung, and Google, dominate both the US and UK samples, together accounting for 87% in the USA and 80% in the UK. Usage of each brand in USA vs. UK is similar (ratio within 0.8 to 1.23) for the top three brands and very different for the other brands. Analyzing cost vs. benefit of calibrating phones, the cost is the 6 minutes to calibrate a phone twice (one model), and the benefit is extending support to a larger fraction of potential participants. The proportion of participants per model is highest for Apple (1.5%), less than half that for Samsung (0.6%) and Google (0.5%), and roughly half that (0.3%) for the other brands. [Disclaimer: another four hundred participants did not finish the study for unknown reasons, which might have biased our sample. We are investigating, and plan to re-run the survey.]

Prolific.com policy requires that participants be informed of their eligibility before they commit to do a study. Participants may not know the model of their phone, but they do know the brand. Thus growth of the profile library emphasizes testing all the models of one brand before proceeding to the next, beginning with the most popular, so Apple, then Samsung, then Google. EasyEyes provides text to the scientist, for inclusion in the Description, that specifies the study requirements, including, when appropriate, the need for a phone of one of several brands.

## B. Computer-model survey

The USA survey also identified the computers used by 199 participants. 74% of the computers ran Windows, 22% macOS, 2% Linux, and 1% ChromeOS. 178 of the 199 computers seem to be unique models, which indicates that the goal of maximizing eligibility of participants for sound calibration is better served by adding microphone than loudspeaker profiles to our library.

## C. 1000 Hz sine calibration

Once the two transducers are selected and identified, the first step is to use a 1 kHz tone to measure the gain at many sound levels. Fig. 6 plots digital power out (dB) versus digital power in (dB) as red circles. The microphone (UMIK-2) was manufacturer-calibrated so the known low-volume microphone gain (-110.5 dB) was used to add a right-hand scale that estimates the sound level (dB SPL). The lower black curve fits a dynamic range compression model (Eqs. 5 & 6) to the data, with an RMS error of 0.1 dB. The parameters of the dynamic range compression model are shown near the top of the figure. The upper part of Fig. 6 shows the Total Harmonic Distortion (THD). The dynamic range compression model is

$$compressed = \begin{cases} T + Q(in - T) & \text{if } T + \frac{W}{2} < in \\ in - \frac{(1-Q)\left(in - \left(T - \frac{W}{2}\right)\right)^2}{2W} & \text{elseif } T - \frac{W}{2} < in \\ in & \text{else} \end{cases} \quad (5)$$

$$out = compressed + gain \quad (6)$$

where *in* and *out* are the digital input and output levels (in dB), *gain* is linear gain (in dB), *T* (in dB) is the compression threshold, *W* (in dB) is the width of transition from linear to full compression, and *Q* is dB-dB slope of the full compression. (Based on Eq. 4 of Giannoulis et al. 2012. Our *Q* replaced their 1/*R*.)

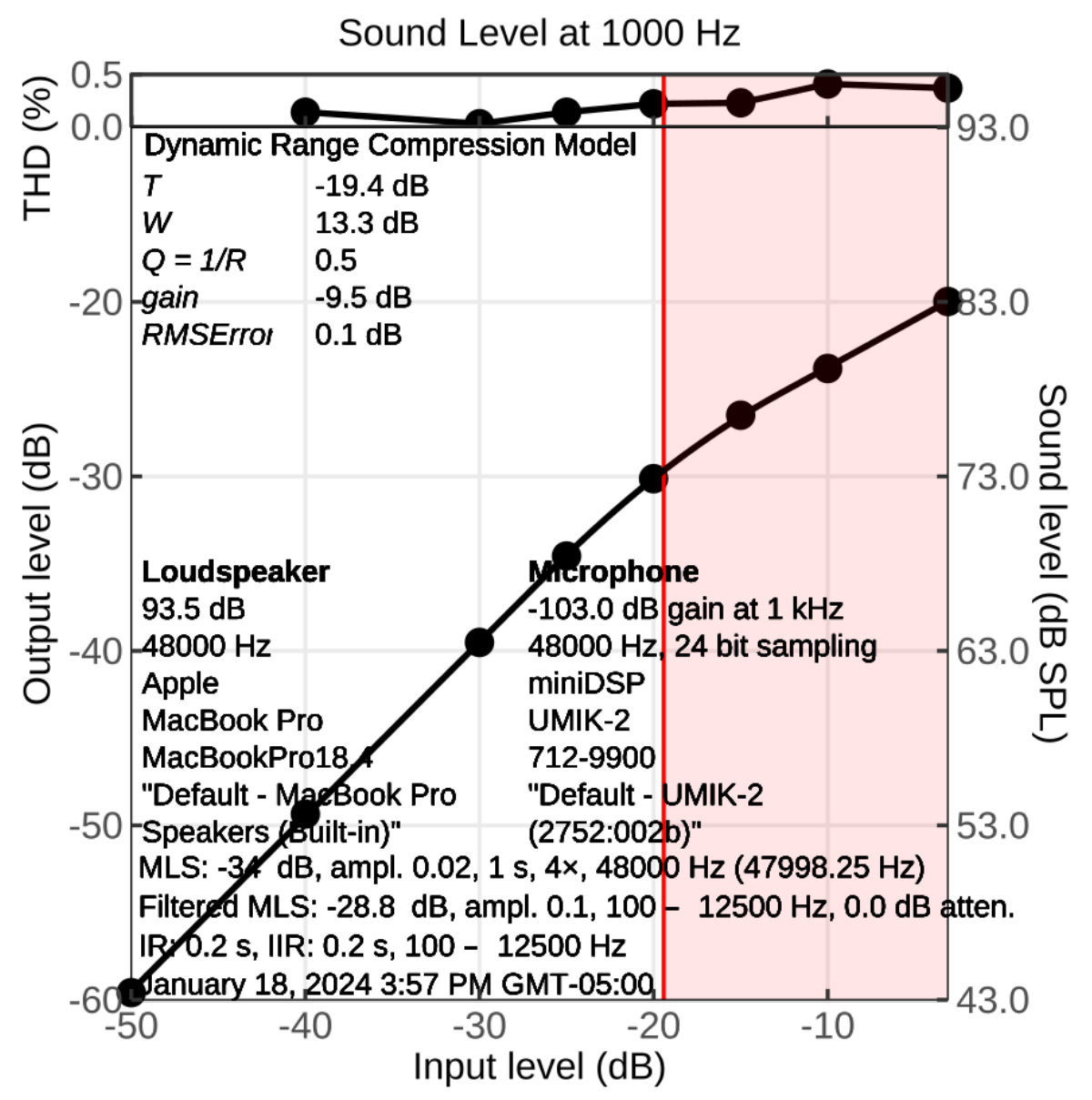

**Figure 6. 1000 Hz sound level**. The lower black curve is the fitted dynamic range compression model (Eqs. 5 & 6), whose parameters are listed in the upper left, with an RMS error of 0.6 dB. The upper black curve is Total Harmonic Distortion (THD). The Loudspeaker and Microphone columns below record essential details of the two transducers. Background sound (and sensor noise in the microphone) are almost entirely rejected by cross-correlating the recording with a 1000 Hz sine to measure 1000 Hz output level uncontaminated by other frequencies. That is why the model fit estimates a negligible additive background level of -70000 dB SPL.

The RMS error (in dB) of the dynamic range compression model fit determines whether the 1000 Hz calibration is accepted. It is repeated if the RMS error exceeds 1 dB. Once accepted, EasyEyes uses only one or two parameters from this fit: *gain* and *T*, both in dB, where *gain* is the gain from digital input to digital output, and *T* is the input threshold of the nonlinearity (in dB). The scientist is allowed to specify the MLS level absolutely or relative to *T*, which is the only use of *T* (calibrateSoundBurstLevelReTBool, default FALSE). In Fig. 6, the vertical red line at input level *T* separates white and pink input ranges. In the white range, the input is less than *T* and the system is nearly linear (log log slope 1 when $in < T - W/2$). In the pink range, the input exceeds *T*, and sound is nonlinearly compressed (log log slope *Q* when $in > T + W/2$, where *Q* is typically 0.5). The compression model curve is inscribed under two straight lines in the log-log coordinates. The two lines meet at $in = T$. The curve is parabolic over a *W*-wide input range centered on $in = T$.

This paper focuses on linear systems analysis of the loudspeaker (and microphone) so it’s important to work in a linear range. Fig. 6 shows that the distortion is higher ($THD \geq 0.2\%$) when the input level exceeds *T* (pink background in Fig. 6), and is lower ($THD \leq 0.1\%$) when it’s below *T* (white background). EasyEyes plays the MLS at a scientist-specified input level to keep distortion down (_calibrateSoundBurstDb, default -34 dB).

*Rejection of background noise*. Calibration is typically done by a scientist at a used phone shop (to calibrate phone microphones) or by a participant at their home, not an anechoic chamber, so background noise is high, typically 40 to 60 dB SPL with a roughly 1/f spectrum. Calibration rejects background noise by cross-correlating the recording with the 1000-Hz-sine or MLS stimulus. The substantial 1/f noise is depicted by the gray curve in Fig. 6, but it is so well rejected by the cross-correlation that the dynamic range compression model (which has no allowance for noise) fits well down to the faintest -50 dB input.

**D. MLS broadband calibration**

As explained in Methods, EasyEyes plays and records an epoch consisting of several, back to back, repeats of an MLS burst. The first period is discarded, to let the audio system settle. The rest are analyzed. Currently the period (_calibrateSoundBurstSec) is 1 second and we analyze four repeats (_calibrateSoundBurstRepeats), so the estimated impulse response is based on 4 seconds of data.

*Correct for Loudspeaker-Microphone difference in sampling rates.* The Novak et al. (2016) nonsynchronous MLS analysis first estimates and corrects for any difference between the loudspeaker and microphone sampling rates. Using a MacBook Pro and a miniDSP UMIK-2 microphone, with a 48000 Hz nominal sampling frequency, across 9 calibrations, the Novak algorithm — by measuring the distance between peaks of the autocorrelation — measured a mean sampling frequency difference of 0.8 Hz with a standard deviation of 0.9 Hz. Novak et al. showed that their algorithm corrects well for even huge 2.18× differences in frequency.

*Three views of the impulse response.* Once the sampling frequency is corrected, the MLS and recording are used to compute the impulse response (Fig. 7). Vanderkooy (1994), based on years of experience using MLS to measure the impulse response of loudspeakers, recommends three diagnostic plots: a 6 ms view of the impulse response, a 50 ms view with 10× higher gain, and a Schroeder (1979) plot, a reverse-time cumulative integral of the impulse response power. All three plots are shown in Fig. 7 (except that Fig. 7B has gain 1× not 10×). Vanderkooy explains how to recognize the spikes and noise-like artifacts that appear in the impulse response when measured in a nonlinear system by the MLS method, typically by playing at a volume exceeding the system's linear range. He notes that the artifacts are largely broad spectrum, including time-specific spikes, and are specific to the particular MLS sequence so that, in principle, one could measure with several different MLS sequences and discard inconsistent features of the impulse response. The Schroeder plot is used to assess reverberation. Ideally it's a straight line, and its slope indicates the reverberation time. Vanderkooy reviews artifacts that can bend the line.

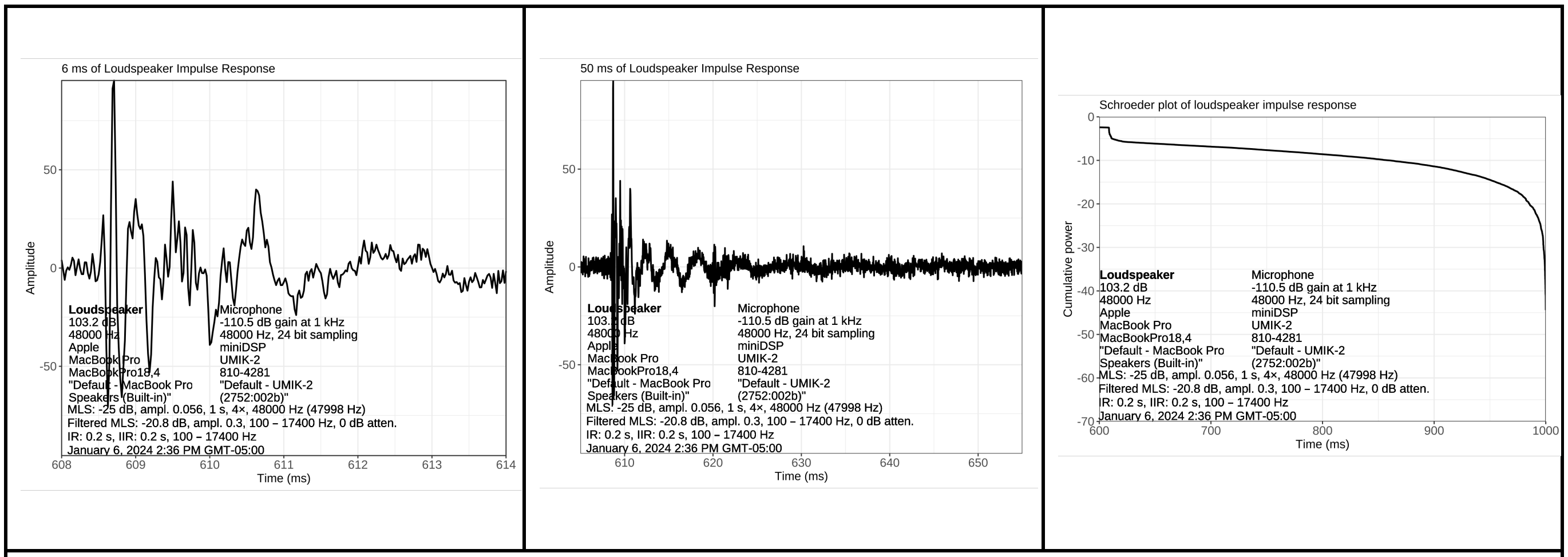

**Figure 7. Impulse response of a 14” MacBook Pro computer’s speakers.** A. 6 ms. B. 50 ms. C. Schroeder (1979) plot, which is the reversed cumulative power over time of loudspeaker impulse response. Reverberation time is usually defined in terms of the slope of the Schroeder curve. The scientist restricted the impulse response length to 200 ms.

*Inverse impulse response*. Fig. 8 provides corresponding displays of the linear-phase inverse impulse response, which is used to correct the loudspeaker. Norcross, Soulodre, & Lavoie (2004) suggest not correcting phase. They found the least physical and perceived distortions using minimum-phase inverse impulse response. EasyEyes currently uses linear phase for the inverse impulse response, and we are considering switching to minimum phase.

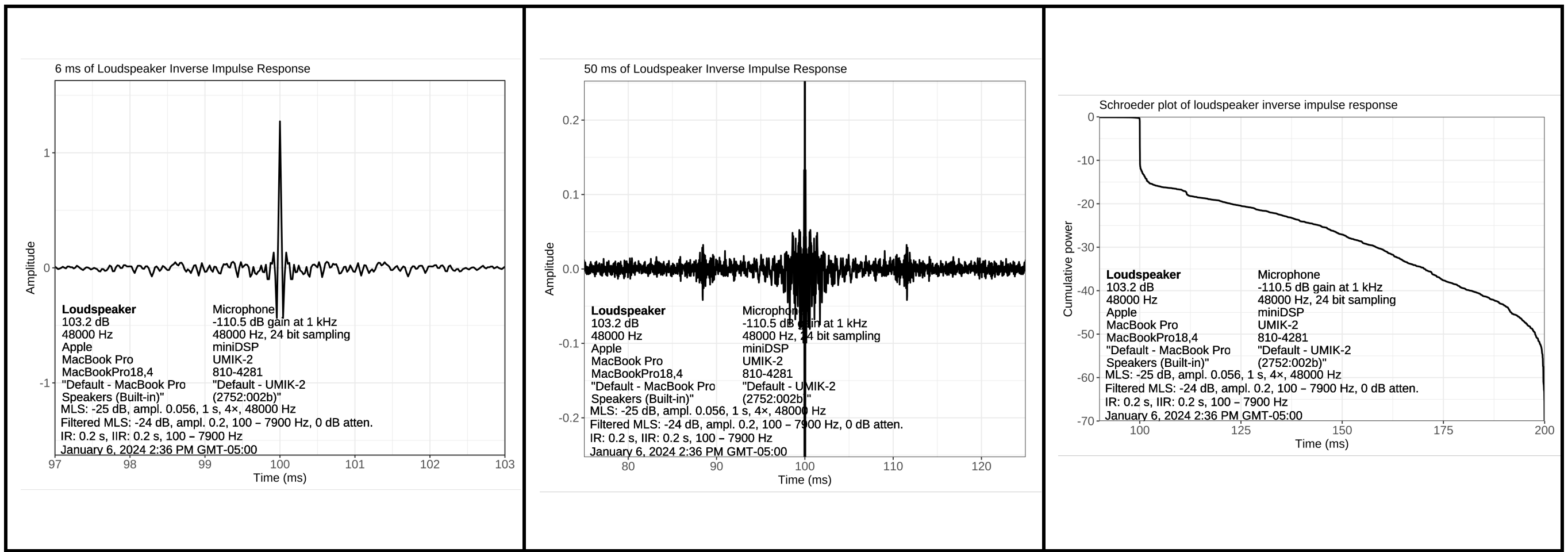

**Figure 8. Linear-phase inverse impulse response used for correction of the loudspeaker in Fig. 7.** A. 6 ms. B. 50 ms and 10╳. C. Schoeder plot. When a source source is corrected by inverse filtering, Norcross, Soulodre, and Lavoie (2004) report that the artifacts of minimum-phase filters tend to be less severe and less audible. The scientist restricted this inverse impulse response length to 200 ms.

**E. Testing the loudspeaker correction**

Typically everything works, and convolving the loudspeaker's audio input with its inverse impulse response (Fig. 8) corrects the loudspeaker to have equal gain at all frequencies up to the scientist-specified upper limit (calibrateSoundMaxHz, default 20000 Hz). Occasionally the MLS procedure gives wildly inaccurate estimates of the impulse response. It's important to exclude the bad estimates, so we assess the quality of each correction. We play an MLS, which has a flat spectrum, through the corrected loudspeaker and check that the sound spectrum is flat.

Our testing has four constraints. First is that, unlike an anechoic chamber, EasyEyes will typically run in ordinary work or living spaces with background noise in the range 40 to 60 dB SPL with a 1/f spectrum. Second, our linear systems analysis requires linearity. Typical laptop loudspeakers aren't very loud, with a max sound level around 85 dB SPL, usually with dynamic range compression that nonlinearly compresses loud sounds, so that the highest linear input sound level (at 1000 Hz) is $T – W/2$. In Fig. 6, for a MacBook Pro, $T – W/2$ is -26 dB, and produces a sound level of 68 dB SPL. We allow the scientist to specify an input power limit (calibrateSoundBurstDb+calibrateSoundBurstFilteredExtraDb) for the correction test, which is currently set to -29 dB. (The scientist can, instead, specify the power limit relative to $T$ by setting calibrateSoundBurstLevelReTBool=TRUE.) Third, compensating for low loudspeaker gain at high frequencies may require a lot of power. Finally, to simplify visual inspection of the spectra, we set the gain so that the IIR filter has gain 1 at 1000 Hz so the filtered MLS has the same power spectral density at 1000 Hz as the original MLS. To satisfy all four constraints, EasyEyes restricts the bandwidth of the filtered MLS.

EasyEyes limits bandwidth to respect the specified power limit. It computes the cumulative integral of the filtered MLS and picks the cut off frequency that gives us a total power within the specified limit, but never exceeding the scientist-specified upper limit (calibrateSoundMaxHz, currently 20000 Hz). We also apply a low-frequency limit (calibrateSoundMinHz, currently 100 Hz) because the 1/f background noise makes it increasingly difficult to measure the signal spectrum at low frequencies.

*Curve by curve.* The loudspeaker correction test is shown in Fig. 9. The story goes from curve to curve. The bottom curve (dashed black) is the profile of the UMIK-2 microphone, provided by the manufacturer. Because this plot is focused on the loudspeaker, the frequency transfer of the microphone is discounted from every plotted recording, so the plots of recordings represent the *sound* spectrum. The dashed red curve is the spectrum of the digital MLS signal sent to the loudspeaker. It has a white, i.e. flat, spectrum. Except for a scalar setting its amplitude, each MLS sample is ±1, so its spectrum is flat up to nearly half the 48000 Hz sampling frequency. The solid red line shows the sound spectrum produced by playing the MLS. It’s not flat. It gradually rises 15 dB from 20 to 200 Hz, has an 8 dB dip at 1400 Hz and lots of ripples in the range 1 to 10 kHz. To correct the loudspeaker, we digitally filter all digital sounds with the inverse impulse response (Fig. 8). The dashed blue curve is the spectrum of the MLS filtered in that way. Notice that it’s been bandlimited to the range 100 to 7900 Hz. The low end is specified by the scientist (calibrateSoundMinHz). The high end is determined by the power limit (calibrateSoundBurstDb+calibrateSoundBurstFilteredExtraDb). We hope the sound spectrum produced by the filtered MLS is flat over the passband. Notice that the dashed blue curve mirrors the red curve. The red curve’s valley at 1400 Hz is matched by a blue peak, and so on. To see this more precisely the dB values of the solid red and dashed blue curves were added to produce the dashed purple curve. The gray curve plots the spectrum of the background noise, recorded in a separate interval while the computer’s speaker was silent. The red curve incorporates ambient background sound along with the MLS-driven sound. To compute the purple curve we first subtract the noise power spectrum from the red curve’s power spectrum.

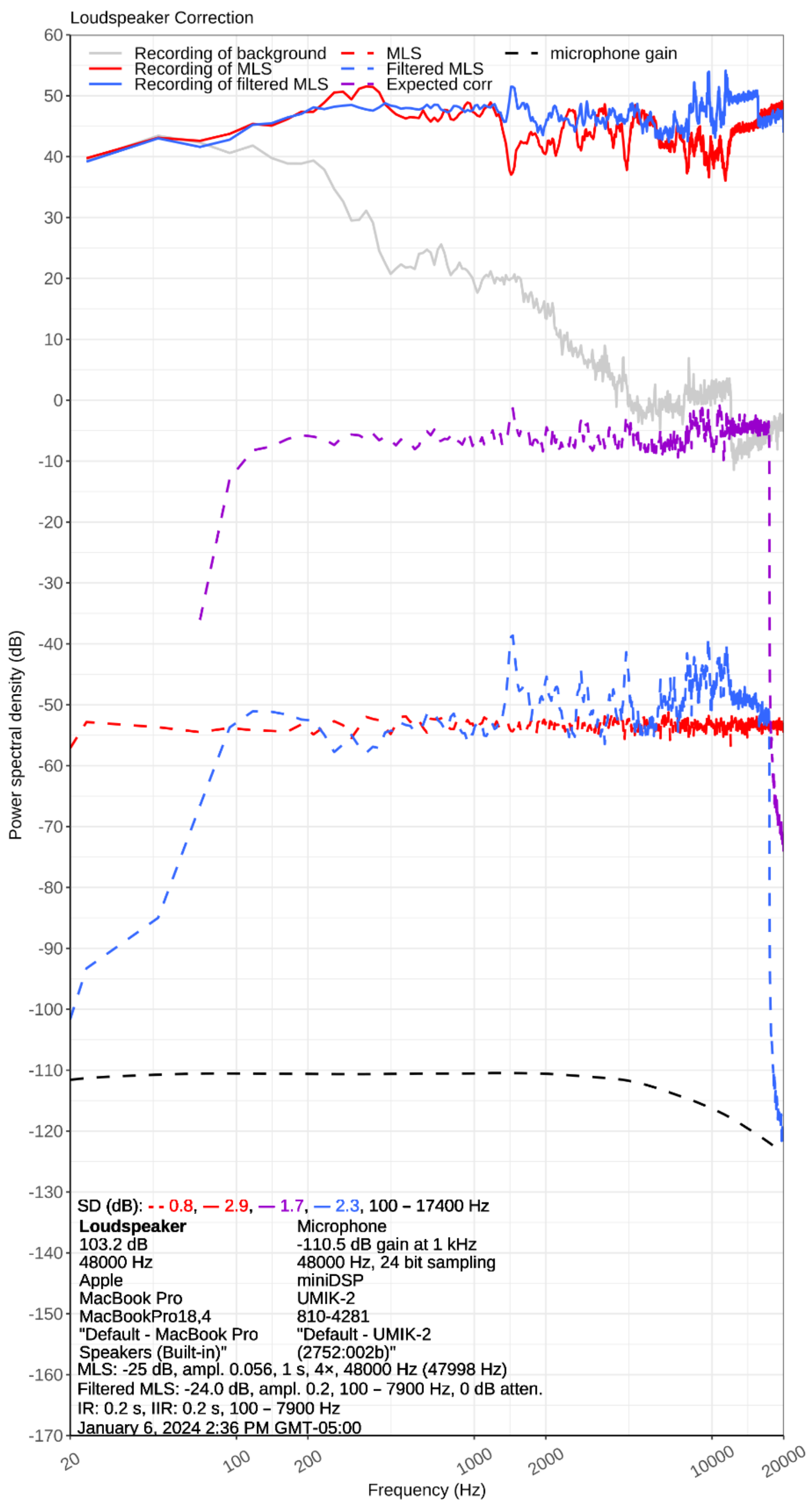

**Figure 9. Assess loudspeaker correction**. Each curve is a power spectrum. Red for the raw MLS. Blue for the filtered MLS, where filtering is convolution with the loudspeaker's inverse impulse response (FIg. R5). The goal of correcting the speaker is assessed by flatness of MLS played with correction, i.e. the 2.3 dB standard deviation of the blue curve over the test range 100 to 9500 Hz. Dashed red and blue are for the digital input to the loudspeaker, and solid red and blue are the sound. The SD (in dB) of these curves, over the tested range of 100 to 9500 Hz, assesses flatness. The red original spectrum is quite flat (sd 1.3 dB) and the blue corrected spectrum is less flat (sd 2.3 dB) by this measure. The gray curve plots the background sound.The dashed purple curve is a sum of dB values to check how well the filtering is expected to correct the loudspeaker. It is the sum of dB values of the dashed blue curve (digital filtered MLS) and the solid red curve (MLS sound), after subtracting out the power of the background noise. If the correction were perfect we'd expect the purple SD to be 0 dB. In fact it's 1.3 dB.

The dashed black line is the frequency response of the measurement microphone. The sound is recorded through the microphone and then corrected for the microphone's frequency transfer function, dashed black curve.

In order to keep the power of the filtered MLS within the linear range, a cut off frequency is imposed, custom for each calibration. Here the test band is 100 to 9500 Hz.

When the noise is bigger than the signal this often produces negative values of power that become NaNs (not a number) when converted to dB. The noise spectrum is roughly 1/f so noise is strongest at low frequencies and prevents us from plotting the purple curve there. Finally the dB values of the dashed blue curve are added to the dB values of the noise-corrected red curve to produce the purple curve. As you can see it's much flatter than either contributing curve, as the wiggles in the dashed blue curve (created by the inverse impulse response) cancel most of the wiggles in the red curve. The solid blue curve shows the sound spectrum produced by playing the filtered MLS. It is quite flat, like the dashed purple curve that predicted it. With perfect cancellation the purple curve would be flat, but in fact it has fine wiggles from 1 to 7 kHz. They are errors in the MLS method, presumably due to slight nonlinearity. Similarly, the sound produced by the filtered MLS (solid blue curve) is not quite flat, having fine wiggles like those in the dashed purple curve.

*Measuring success.* The goal is to correct the speaker, so we measure how well we succeeded by computing the standard deviation of the blue spectrum over its passband (here 100 to 7000 Hz). Note that the plot gives equal space to each decade, 20 to 200, 200 to 2000, and 2000 to 20000 Hz, but the computation of standard deviation uses the spectrum produced by the FFT which has samples equally spaced in frequency, so the second and third decades contribute ten and hundred times more samples to the computed standard deviation. We reject the loudspeaker calibration if the standard deviation (over its bandwidth) of the filtered MLS sound spectrum exceeds 3 dB. If it's rejected then we measure it again. Roughly a third of the calibrations are rejected in this way.

**F. Repeated calibration**

A key metric of data quality is the repeatability of measurement. Microphones can be calibrated by any scientist with a measurement microphone. Fig.10 shows 8 profiles of the same iPhone 12 Pro made one after another. The EasyEyes profile library consists of profiles created in this way. First a measurement microphone is used to calibrate the computer's loudspeaker. Then the loudspeaker is used to calibrate microphones. It takes three minutes per calibration. In Fig.10 the standard deviation at 1000 Hz is 1.4 dB. It's clearly higher for the lowest frequencies (20 to 100 Hz) but those frequencies are rarely used in hearing studies.

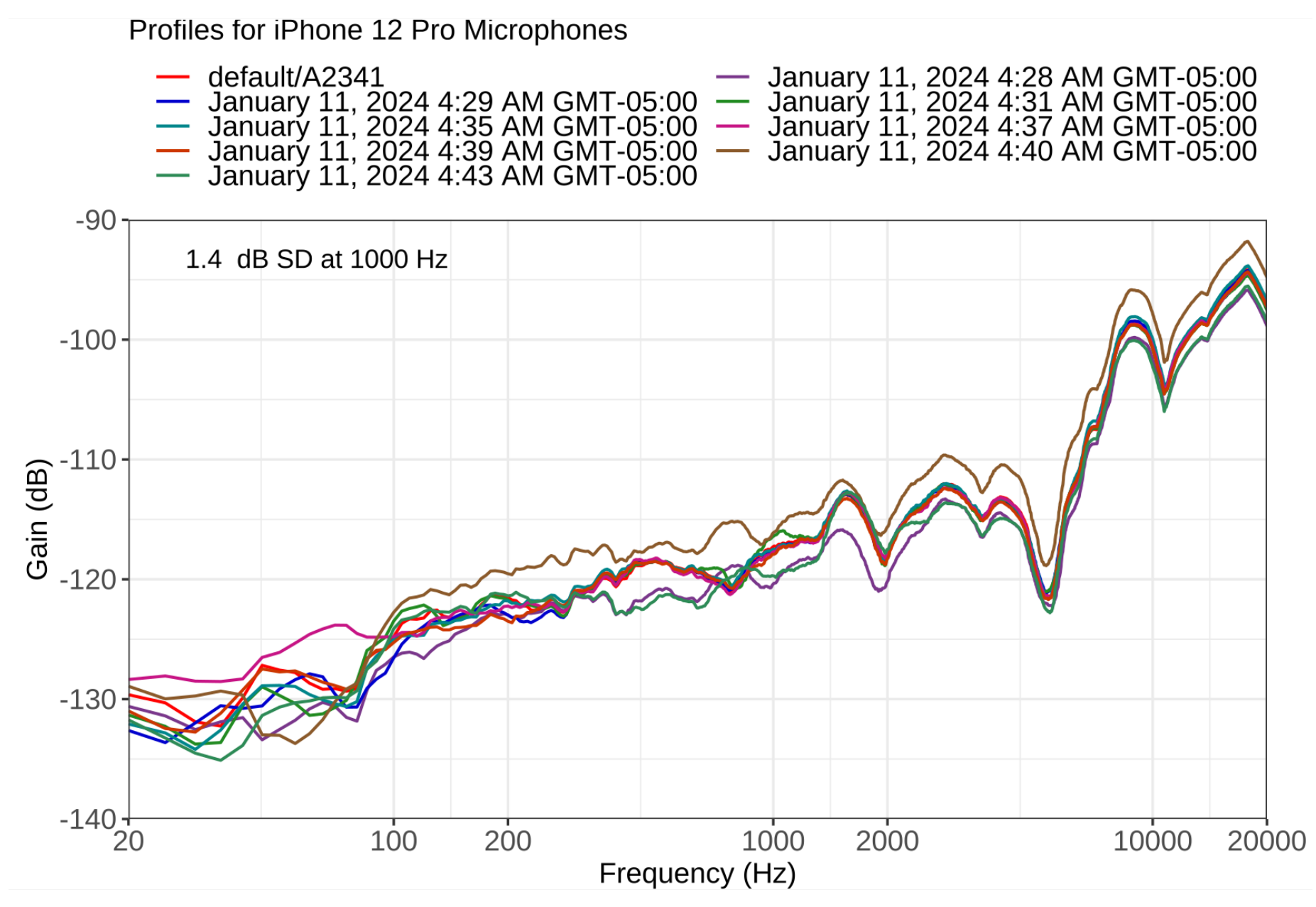

**Figure 10. iPhone 12 Pro microphone** (A2341): 8 (of 9) profiles (correctionSD ≤ 2.5 dB) calibrated by MacBook Pro speaker (MacBookPro18,4) calibrated by UMIK-2 (810-4281).

correctionSD = 1.2±0.1 (mean±SD)
1000 Hz gain = -117.8 ±1.4 dB
T = -19.5±2.0 dB
W = 12.4±6.0 dB
Q = 0.5±0.1
modelRMSError = 0.3±0.2 dB
fs2 = 48001.6±0.6 Hz

Each participant will use their phone, with its matching library profile, to calibrate their computer’s speakers. Fig. 11 shows nine profiles for the speakers of one MacBook Pro, repeatedly calibrated by the same iPhone 12 Pro. The standard deviation is 0.6 dB at 1000 Hz.

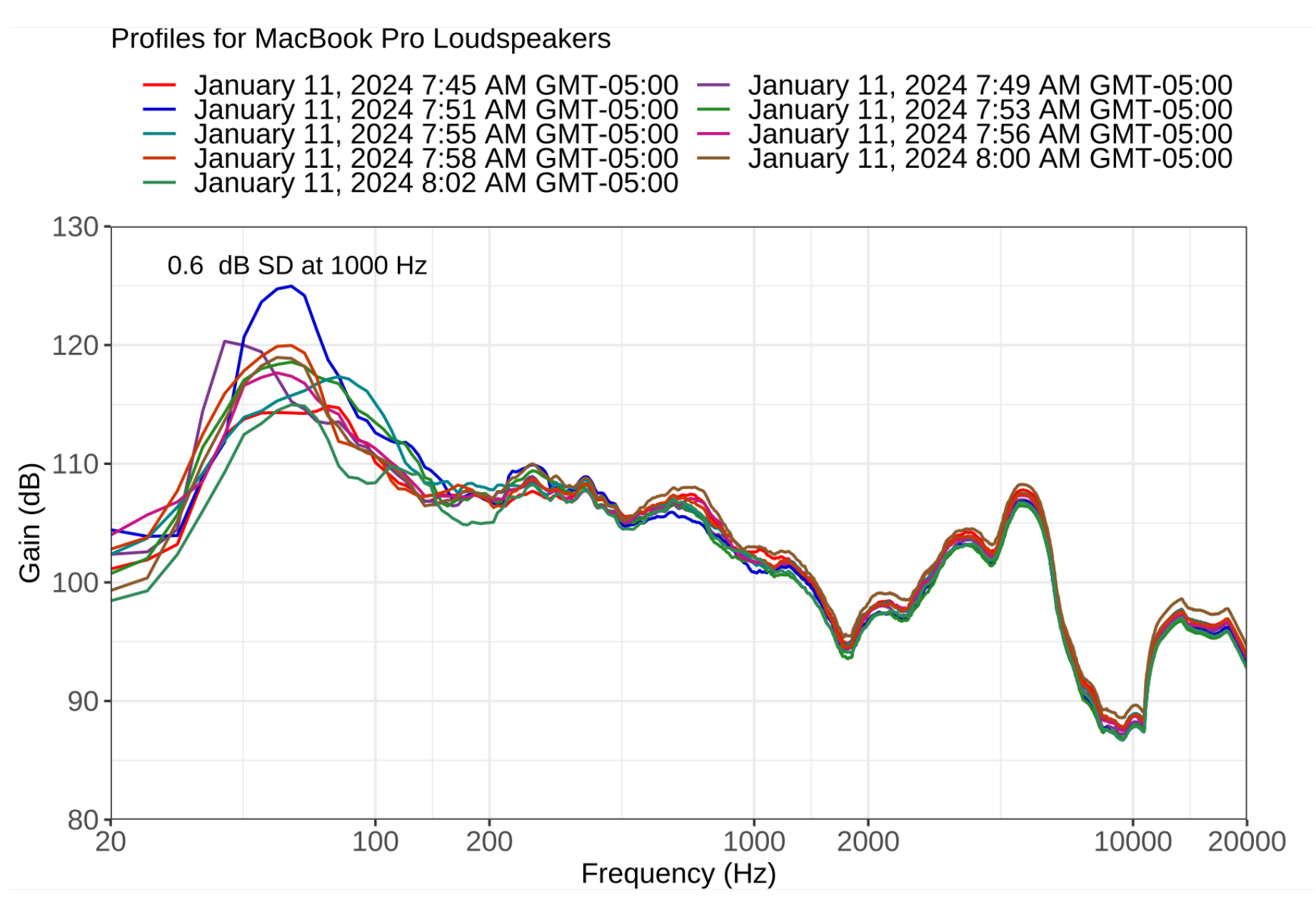

**Figure 11. MacBook Pro speakers** (MacBookPro18,4): 9 (of 10) profiles (with correctionSD ≤ 2.5 dB) calibrated by iPhone 12 Pro using library profile.

correctionSD = 1.3±0.1 dB (mean±SD)
1000 Hz gain = 101.9±0.6 dB
T = -19.9±5.3 dB
W = 11±5 dB
Q = 0.6±0.1
fs2 = 48001.4±0.6 Hz

### G. Documenting the calibration

Each calibration provides an 8 MB JSON file to the scientist documenting the calibration. Shiny is a web interface for R graphics. The EasyEyes “Analyze: Sound” Shiny web page reads the JSON file and produced Figs. 4, 5, 6, 7, 8, and 9. The EasyEyes “Analyze: Profiles” Shiny web page is a front end to the profile library and produced Figs. 10 and 11. Currently the figures can be downloaded in several formats (pdf, png, eps, svg). We hope to add the option of a Jupyter notebook with R code.

## IV. DISCUSSION

Hearing researchers are accustomed to working with sounds that are accurate to 1 dB. Our repeated testing indicates that each step of the calibration chain and a standard deviation of about 1.5 dB. The three calibrations, from measurement microphone to computer speaker to phone microphone (saved in library) to participant computer's speaker, will contribute independent errors, and thus a total standard deviation of $\sqrt{3}\times1.5 = 2.6$ dB. That's high for hearing research, but we have straightforward ways to improve it.

*Reducing noise.* The multiple profiles shown in Figs. 10 and 11 seem to show independent variance at each frequency, which suggests that it comes directly from the MLS measurement. The signal to noise ratio of that technique improves with duration of the MLS period and the number of periods recorded (Novak et al. 2016). In the current 3-minute calibration, EasyEyes analyzes only 4 seconds (4 repeats of a 1 second period) of recorded unfiltered MLS, so we could increase that recording length tenfold with only a modest effect on the whole calibration time.

*Reducing artifacts.* As noted above, Vanderkooy (1994) observes that the artifacts of the MLS method in the measured impulse response are mostly brief broad-spectrum spikes at times unique to the MLS sequence. He suggested that one could measure the impulse response with several different MLS sequences and edit out the artifacts. It would be easy to enhance our software to measure the impulse response three times using different MLS sequences. Following Vanderkooy's observation, using a point-by-point median to combine the three impulse responses might be effective in removing MLS-method artifacts.

*Individual differences.* Beyond the standard deviation of repeated measurement, there will also be variations between phones of the same model (which we hope will be small), and variations in phone placement across participants. We will assess these, and anticipate that they can be managed, possibly requiring improved instructions, to be less than the SD of repeated testing.

*Low-frequency gain of laptop loudspeakers.* The low-frequency gain of the MacBook Pro speaker in Fig. 11 is remarkably good and might seem too good. However, Salman (2022) reported that the then-new 2021 16-Inch MacBook Pro and 2022 17-inch Razer Blade Pro loudspeakers have much better low-frequency gain than all other laptops that he tested. In his report, all the loudspeaker frequency transfer functions roll off at low frequencies with a log-log slope in the range 2 to 6. That roll off begins at about 100 Hz for the two best laptops, and at 200 Hz, or more, for the rest.

Current developments:

1. Rerun the phone survey because we had only a 50% retention rate, apparently due to a runtime bug.
2. Use a measurement microphone to measure RMS error of sound correction. This software is nearly working.
3. Increase the MLS period duration or number of repetitions to reduce profile variance.
4. Inspired by Vanderkooy (1994), remove MLS artifacts by taking the point-by-point median of impulse responses measured with three different MLS sequences.
5. Switch from linear phase to minimum phase inverse impulse response, as recommended by Norcross, Soulodre, & Lavoie (2004).
6. Build the phone profile library: Apple, then Samsung, then Google.
7. As we build the library, we will test the assumption that two phones with the same model name have identical microphone acoustics by testing at least two phones of each model.
8. Measure variance of the loudspeaker profile for a particular computer model across participants, each calibrating with their own phone.
9. Offer a Jupyter notebook with R code as an option for downloading plots from the Shiny pages that produced Figs. 4 – 11.

## V. CONCLUSION

To go online, hearing research must provide calibrated sound to online participants. Here we showed that a library of smartphone-microphone profiles enables use of the participant's phone to calibrate their computer's loudspeaker in approximately three minutes using the Novak et al. (2016) nonsynchronous MLS algorithm. The participant's phone model is selected by the participant and verified by its screen size. Currently the resulting loudspeaker calibrations have a standard deviation of 3 dB across a frequency range of 100 to 20,000 Hz, based on a 4-second recording. Recording longer will improve this precision. Additionally, Vanderkooy (1994) suggested that combining recordings with different MLS sequences might dramatically reduce MLS artifacts. Our survey of 400 Prolific.com participants shows that supporting three phone brands is enough to support 87% of potential participants in the USA and 80% in the UK. In summary, this online loudspeaker-calibration method allows hearing research to go online.

## SUPPLEMENTARY MATERIAL

EasyEyes is open source, available at github.com/EasyEyes. This includes our Python impulse-response nonsynchronous measurement code, based on the Novak et al. (2016) MATLAB code. The profile library is accessed through a front end that is part of the easyeyes.app web site.

## ACKNOWLEDGEMENTS

We thank Antonin Novak for sharing the MATLAB source code developed by Novak et al. (2016) to measure the impulse response using nonsynchronous loudspeaker and microphone. We translated it to Python. We thank Antje Ihlefeld for pointing us in the right direction, Mario Svirsky for advice on sound calibration, Sarah Master for proofreading, and Laura Suciu for helpful suggestions. Supported by NIH core vision grant P30-EY013079.

## AUTHOR DECLARATIONS

The authors have no conflicts of interest. It is planned that New York University will charge scientists a small fee per participant tested for use of the open-source EasyEyes.com platform. This will support maintenance and development of the software. Denis Pelli will manage that but will not profit from it.

## AUTHOR CONTRIBUTIONS

Ivan Vican wrote the second draft of this paper, and acted as auditory signal processing consultant to the project. Hugo De Moraes translated the Novak et al. (2016) MATLAB MLS signal processing code to Python, integrated it into the mostly JavaScript EasyEyes, and wrote the first draft of this paper. “Simon” Chongjun Liao implemented the Shiny R plots in the EasyEyes Analyze: Sound and Profiles web pages, and also contributed to the Python signal processing. Nathnael Tsegaya implemented the profile library and the Needs page. William O’Gara enhanced the Python signal processing software. Denis Pelli conceived, supervised, and tested the project, and wrote this third draft.

## DATA AVAILABILITY

We plan to provide a Jupyter notebook in R for each of the figures with data (Figs. 4 to 11).